%
\input mn.tex
\input epsf.tex

\hfuzz=20pt
\def \ten{\times 10}
 
\def \flux{\hbox{erg}\,\hbox{cm}^{-2}\,\hbox{s}^{-1}}
\def \lum {\hbox{erg}\,\hbox{s}^{-1}}
\def \rlum{\hbox{mJy}\,\hbox{kpc}^2}
\def \gam{$\gamma$}

\begintopmatter 

\title{The implications of radio-quiet neutron stars}

\author{K.T.S. Brazier$^1$ and S. Johnston$^2$}

\affiliation{$^1$ Dept. of Physics, University of Durham,
             South Road, Durham DH1 3LE}
\vskip 3pt
\affiliation{$^2$ Research Centre for Theoretical Astrophysics,
		University of Sydney, NSW 2006, Australia}

\acceptedline{Accepted   .............. Received   ..........   \today.}

\abstract 
{ We collate the evidence for rotation-powered neutron stars that are
visible as X--ray sources and not as radio pulsars.  To date, ten objects
have been proposed and one, Geminga, has been confirmed as a pulsar by
the detection of 4.2 Hz pulsations.  Several indicators have been used
to support the proposition that the X--ray sources are isolated neutron
stars, including high X--ray to optical/radio flux ratios, a constant
X--ray flux and coincidence with a \gam--ray source.  Seven of the
published neutron star candidates are located near the centres of
supernova remnants, two of them within plerions, suggesting that these
are young objects ($\tau < $20,000 yr).  The remaining candidate neutron
stars have no associated supernova remnant and may be older systems,
powered either by their rotation, like Geminga, or possibly by accretion
from the interstellar medium.

Quantitative upper limits exist for the radio fluxes of eight of the ten
objects and reveal a population at least an order of magnitude less
luminous at radio wavelengths than known radio pulsars of similar power
or age.  These could be intrinsically low
luminosity pulsars, but this implies an overpopulation of neutron stars
relative to the galactic supernova rate. A simple alternative
explanation within the context of existing pulsar models is that these
objects are pulsars in which the radio beams are directed away from
Earth.  They are still visible as X--ray sources because the weakly
modulated, surface (thermal) emission, which dominates the soft X--ray
emission in most young to middle-aged radio pulsars, is radiated in all
directions.  In the cases where hard X--ray or \gam--ray fluxes are
seen, the beaming explanation implies different emission sites for the
non-thermal high-energy radiation and the unseen radio beams.  From the
number of candidate neutron stars and radio pulsars younger than 20,000
years and within 3.5 kpc, the radio beaming fraction of young pulsars is
estimated to be roughly 50\% and certainly much less than 100\%.  We
find the local neutron star birth rate to be at least 13
Myr$^{-1}\,\hbox{kpc}^{-2}$.  This extrapolates to a galactic rate of
one neutron star born every $\sim 90$ years.  We conclude that probably
all neutron stars are born as radio pulsars and that most young, nearby
pulsars have already been discovered.  
}


\maketitle

\section{Introduction}

The Princeton pulsar catalogue (Taylor et al. 1993) now contains entries
for over 700 radio pulsars, collected from many pulsar surveys and
targeted searches.  In addition to the radio pulsars, there is a single
entry with no measured 400 MHz or 1400 MHz flux: the X-- and \gam--ray
pulsar known as Geminga (Halpern \& Holt 1992, Bertsch et al. 1992).
The tight upper limits on its radio flux give Geminga a luminosity limit
several orders of magnitude below those seen in known radio pulsars
of similar age or power and place this source apart from the general
population.

At high energies, there is no obvious difference between Geminga and the
$\sim 20$ other pulsars in the Princeton catalogue that have been
detected as X--ray sources.  Seven, including Geminga, have also been
seen by \gam--ray telescopes.  The high energy detections are limited to
the very brightest objects and sample a more powerful and nearby set of
pulsars than the deeper radio searches.  In X--rays, the luminosity is
approximately correlated with pulsar spin-down power, $\dot E$, so that
$\dot E$ divided by the square of the distance $d$ is a convenient
measure of the detectability of a radio pulsar as an X--ray source.  The
radio pulsars detected in the ROSAT (0.1--2.4 keV) band all have $\dot E
/ d^2 \ga 10^{34}\,\hbox{erg}\,\hbox{s}^{-1}\hbox{kpc}^{-2}$ and,
conversely, nearly all of the radio pulsars fitting this criterion have
been detected.  It is important to note that the detections include not
only young, powerful pulsars, but also old and millisecond pulsars that
are less powerful but are nearby.  Taking distance uncertainties into
account, the high level of X--ray detections therefore implies that most
radio pulsars are X--ray pulsars.  The inverse, that most or all X--ray
pulsars are radio pulsars, need not be true: Geminga is a specific
counter-example.  Note that the term `X--ray pulsar' is taken here and
throughout this paper to mean a rotation-powered pulsar observed in
X--rays, {\it not} a neutron star powered by accretion from a stellar
companion or from a residual accretion disk.

Geminga is a 350,000-year-old pulsar at a distance of only $\sim
160\,$pc.  Its proximity has allowed very low luminosity limits to be
calculated from the limits on its radio flux and it has been labelled
`radio-quiet'. However, a low flux density does not necessarily mean
that the radio {\it luminosity} is low.  It could simply mean that the radio
beam is not visible from the Earth.  In this paper we will define a
radio-quiet pulsar as a rotation-powered pulsar which has not been
detected at 400 or 1400 MHz and which therefore has a low {\it inferred}
luminosity at these frequencies.  This definition does not distinguish
between explanations based on luminosity and beaming, nor does it equate
radio-{\it quiet} with radio-{\it silent}.

Radio-quiet pulsars are extremely difficult to identify because photon
statistics at high energies make useful pulsation searches impossible
for all but the brightest sources.  The first evidence for a radio-quiet
pulsar is usually an extreme spectrum: bright in X--rays and/or \gam--rays
but very faint at optical wavelengths.  While Geminga is the sole
confirmed radio-quiet pulsar, nine further objects have been discovered
as unresolved X--ray sources and proposed as radio-quiet,
rotation-powered neutron stars after fulfilling the first criteria for
detection of a pulsar.  In this paper, we add these to Geminga to examine
the case for radio-quiet pulsars as a class and to distinguish between
beaming and luminosity explanations for these objects.

\begintable*{1}
\tabletext{
{\bf Table 1.}  Candidate isolated neutron stars and their properties.  
X--ray fluxes are from ROSAT and are used to derive a
spin-down power $\dot E$ except where an alternative estimate is listed.
These cases are discussed in the text.  Radio flux limits are as given
in the references and are a mixture of pulse and point source limits
as available.  Distances and ages are for the supernova remnants, where 
one has been associated with the neutron star candidate.  References:  
1.  Slane et al. 1997;
2.  Brazier et al. 1997;
3.  Helfand et al. 1995;
4.  Petre et al. 1996;
5.  Walter et al. 1996;
6.  Neuh\"auser et al. 1997;
7.  Frail \& Moffett 1993;
8.  Seiradakis 1992;
9.  Hailey \& Craig 1995;
10. Brazier et al. 1996;
11. Tuohy \& Garmire 1980;
12. Gotthelf et al. 1997;
13. Stocke et al. 1995;
14. Kaspi et al. 1996;
15. Mereghetti et al. 1996;
16. Lorimer et al. 1997b.
}
\halign {$\hbox{#}\hfil$&&\quad$#\hfil$\cr
\noalign{\hrule\vskip 0.03in}\cr
\noalign{\hrule\vskip 0.08in}
Name		&F_x(0.1-2.4\,\hbox{keV})&\hfil\dot E	&\hfil\hbox{S}_{400}&\hbox{S}_{1400}    &\hbox{SNR}	&\hbox{Common} &\hfil\hbox{dist}  &\hfil\hbox{age}&\hbox{refs}\cr
{}		&(\flux)		&(10^{30}\,\lum)&\hbox{(mJy)}	    &\hbox{(mJy)}	&\hbox{name}    &\hbox{name}  &\hbox{(kpc)} &(10^3 \hbox{yr}) &{}    \cr
\noalign{\vskip 0.03in\hrule\vskip 0.08in}

RXJ0002+6246	&\quad 2\ten^{-13}	&		&\quad ?	&\quad\quad ?	&\hbox{G117.7+0.6}	&{}		&3	     &20	       &9\cr
RXJ0007.0+7302	&\quad 9\ten^{-14}	&7\ten^6	&<1.5 		&\quad <0.3	&\hbox{G119.5+10.2}	&\hbox{CTA1}	&1.4	     &5-10	       &1,2,16\cr
RXJ0201.8+6435	&\quad\quad ?		&(2-5)\ten^6	&<2.1		&\quad <0.15	&\hbox{G130.7+3.1}	&\hbox{3C58}	&3.2	     &0.8	       &3,7,16\cr
1E 0820--4247	&\quad 3\ten^{-12}	&		&<1.7		&\quad <0.3	&\hbox{G260.4-3.4}	&\hbox{Pup A}	&2	     &3.7	       &4,14\cr
1E 1207.4--5209	&\quad 2\ten^{-12}	&		&<1.0		&		&\hbox{G296.5+10.0}	&\hbox{PKS 1209-51/52}	        &1.5	      &7               &14,15\cr
1E 161348--5055	&\quad 7\ten^{-13}	&		&<1.5		&\quad <0.1	&\hbox{G332.4-0.4}	&\hbox{RCW103}	&3.3	     &1-3              &11,12,14\cr
RXJ2020.2+4026	&\quad 4\ten^{-14}	&7\ten^6	&<5		&\quad <0.35	&\hbox{G078.2+2.1}	&\gamma\hbox{--Cyg}&1.5	     &10	       &10,16\cr
\noalign{\vskip 0.03in\hrule\vskip 0.08in}
\noalign{\hbox {-- cases with no SNR --}}
Geminga		&\quad 4\ten^{-13}	&3\ten^4	&<0.1		&\quad <1.0	&	&	&0.16        &350              &8\cr
RXJ1856.5--3754&\quad 1\ten^{-11}	&		& <4		&		&	&	&0.12	     &	               &5,6\cr
MS 0317--6647	&\quad 4\ten^{-13}	&		& \quad ?	&\quad \quad?	&	&	&	     &	               &13\cr
\noalign{\vskip 0.03in\hrule\vskip 0.08in}
}
\endtable

\section{The candidate pulsars}

Nine candidate pulsars are listed in Table 1, together with Geminga.
Each of the candidates is an unresolved X--ray source that has not been
associated with a compact radio object, leading to the proposal that
they are rotation-powered, radio-quiet neutron stars.  Seven of the
objects are within supernova remnants.  Beyond these generalities, the
details diverge of the sources and the work done to identify them.  The
table does not include the `anomalous X--ray pulsars' discovered in SNRs
-- 1E2259+586, RXJ1838.4--0301 and 1E1841--045 (Gregory \& Fahlman 1980,
Schwentker 1994, Vasisht \& Gotthelf 1997) -- which are thought to be
isolated but powered by accretion from a residual disk (van Paradijs et
al.  1995).

The X--ray fluxes listed in Table 1 provide a useful way to estimate the
spin-down power $\dot E$ of the candidate pulsars.  Older radio pulsars
detected as point X--ray sources follow an approximate trend of $L_x \sim
3\times 10^{-4} \dot E$ (cf. Seward \& Wang 1988, \"Ogelman 1995) which
we can invert to estimate the spin-down power from the X--ray luminosity.
Since the trend includes only the detected X--ray pulsars, it is biased
towards high X--ray luminosities and an $\dot E$ derived from it may be
underestimated.  No allowance for the bias is made in this paper because
the candidate pulsars are also selected as X--ray sources.

For 3C58 and CTA 1, the plerions provide an independent way to estimate
the power of the embedded pulsars.  Seward \& Wang (1988) found an
empirical relationship between the X--ray luminosity of a plerion and the
spin-down power of the pulsar powering the plerion.  For CTA 1, Slane et
al. (1997) used this relationship to estimate a spin-down power of $1.7
\ten^{36} \lum$ .  For 3C58, similar considerations of the X--ray
synchrotron emission yield $\dot E = (2\hbox{--}4) \ten^{36} \lum$
(Helfand et al. 1995), while $\dot E = 1.5 \ten^{36} \lum$ is required
to explain a radio filament near to the candidate pulsar in terms of a
shock in the pulsar wind (Frail \& Moffett 1993).

The pulsar candidates in CTA 1 and G078.2+2.1 are coincident with
\gam--ray sources, providing a third way to estimate the spin-down
power.  While the relationship between \gam--ray luminosity and
spin-down power is still unclear (Fierro 1995), Brazier et al. (1996,
1997) showed that the \gam--ray luminosities inferred for the two
candidates were consistent with those of Vela and PSR B1706--44.  In both
cases, the spin-down power derived from the X--ray flux is more than an
order of magnitude lower than the \gam--ray comparison would suggest.
The estimates from the \gam--ray fluxes are listed in Table 1.  An
intermediate spin down power is used in this paper for these two
objects.

\vskip 15pt 

It would be unwise to treat the candidates in supernova remnants as a
statistical sample.  ROSAT, used to observe each of them,
has not observed many of the Galaxy's SNRs, and identification of a
neutron star candidate depends on the interests of the observer as well
as an object's physical properties.  In addition, the sensitivity to
point sources within an SNR varies with the intensity profile of the SNR
emission, so that objects close to the rim of a SNR may be submerged in
bright diffuse emission.  In RCW 103, for example, the ASCA observation
has confirmed the status of the point source 1E~161358--5055 by
separating it spectrally from the SNR (Gotthelf et al. 1997).

In addition to the candidate pulsars within supernova remnants, there
are a further two candidates, plus the confirmed pulsar Geminga, which
are not associated with SNRs.  The list is short because there is much
less to attract the attention of observers than in the case of an SNR.
Without the bonus of a supernova remnant, it is also difficult to
estimate the distances to these pulsar candidates or to guess their
ages.  Nevertheless, the one confirmed radio-quiet pulsar is in this
group.

\subsection{Candidate pulsars within SNRs}

\subsubsection{RXJ0002+6246 (G117.7+0.6)}

A faint X--ray arc was discovered by Hailey \& Craig (1995) in the
periphery of ROSAT PSPC images of the SNR CTB 1.  The arc was proposed
to form part of a previously unknown shell-type supernova remnant,
G117.7+0.6 [footnote: The proposed remnant is centred at galactic
coordinates $(l,b) = (117.4,0.3)$.  The position $(117.7,0.6)$ appears
to be a mis-conversion from celestial to galactic coordinates.] at a
distance of $\sim 3$ kpc.  An unidentified point source, RXJ0002+6246,
is clearly visible 10 arcmin from the central position.  None of the
nearby optical sources was found to be a plausible counterpart and there
is no radio source listed at this position, suggesting that RXJ0002+6246
could be a radio-quiet neutron star associated with the proposed
remnant, possibly with a pulsation period of 242 ms (Hailey \& Craig
1995).  Because no quantitative limit to the radio flux is given,
RXJ0002+6246 will not be included in this paper as a radio-quiet
pulsar candidate.

\subsubsection{RXJ0007.0+7302 (CTA 1)}

This object lies in the north of the CTA 1 supernova remnant, a mostly
circular radio SNR with a `blow-out' on its north-east side (Pineault et
al. 1993).  In X--rays, the SNR is dominated by non-thermal plerionic
emission, interpreted as synchrotron emission from the relativistic
particles generated by a fast pulsar (Slane et al. 1997).  Slane et
al. propose that the point source RXJ0007.0+7302, which is positioned at
the centre of the plerion, is the pulsar, and this is supported by
limits on its optical and radio flux (Brazier et al. 1997).  The flux
and spectrum of a persistent, unidentified \gam--ray source coincident
with RXJ0007.0+7302 are consistent with emission from a pulsar at the
age and distance of CTA 1, leading Brazier et al. (1997) to propose that
this object, like Geminga, is a \gam--ray loud, radio-quiet pulsar.  If
we accept this proposal, the spectrum peaks in the GeV region, making
this a `Vela-like' rather than `Crab-like' pulsar, with the faint flux
in soft X--rays typical of pulsars older than a few thousand years.

\subsubsection{RXJ0201.8+6435 (3C58)}

3C58 (G130.7+3.1) is a young SNR, similar to the Crab SNR in radio
spectrum and morphology, and is thought to have resulted from the
supernova of AD 1181 (Clark \& Stephenson 1977).  Confirmation of a
compact X--ray source within the plerion (Becker et al. 1982) was
provided by Helfand et al. (1995).  It is still unclear whether a
north-south ellipticity of the X--ray source means that the object is in
fact associated with a radio filament discovered by Frail \& Moffett
(1993) or that the apparent extent is due to the attitude reconstruction
problems known to affect some ROSAT data.  Helfand et al. (1995) determine that
the X--ray luminosity of the plerionic SNR emission is (4--7)$\ten^{33}
\lum$, which would require the neutron star to have a spin-down power of
$\dot E \sim 10^{35}$ -- $4\ten^{36} \lum$, bracketing the estimated
power of $1.5 \ten^{36} \lum$ needed to produce a shock at the position
of the radio filament (Frail \& Moffett 1993).  This is much smaller
than than the power of known radio pulsars with such small ages.  The
ROSAT spectrum and the upper limit of 50\% on pulsations are best
modelled as thermal emission from the hot polar caps of a neutron star,
not non-thermal magnetospheric pulses.  The properties of the nebula and
the compact source can be reconciled within the age constraints if the
pulsar is relatively slow and with a surface magnetic field above
$10^{13}\,$G (Helfand et al. 1995).

\subsubsection{1E 0820--4247 (Puppis A)}

A compact source close to the centre of Puppis A was first discovered in
Einstein HRI images (Petre et al. 1982), but its point-like nature has
only recently been confirmed (Petre et al. 1996).  Stringent optical and
radio limits (Petre et al. 1996, Kaspi et al. 1996) rule out most types
of X--ray sources apart from a neutron star or a BL Lac with weak radio
emission.  The X--ray source shows no evidence of variability between
observations and is less than 20\% pulsed,
implying that, if this is a neutron star, the soft X--ray emission is
predominantly thermal rather than magnetospheric, i.e. more similar to
the Vela pulsar than to the Crab.  There is no visible plerion.

\subsubsection{1E 1207.4--5209 (G296.5+10.0)}

G296.5+10.0 has a symmetric `barrel' morphology defined by two clear,
elongated radio arcs; its age is estimated to be roughly 20,000 years
(Seward \& Wang 1988).  Using Einstein observations of the remnant,
Kellett et al. (1987) showed that a compact X--ray source, 1E
1207.4--5209, close to the geometric centre of the remnant might be an
isolated neutron star.  Matsui et al. (1988) strengthened this claim
with upper limits on the optical luminosity, while Mereghetti et
al. (1996) have published further details of the effort to identify this
X--ray source.  They find no plausible optical counterpart down to $m_V
\sim 25$ and place a 0.1 mJy limit at 4.8 GHz.  
Kaspi et al. (1996) place a tighter 1 mJy pulsation limit at
400 MHz.  Recent analysis of the ASCA/ROSAT spectrum of 1E 1207.4--5209
shows that it is thermal and can be interpreted in terms of a cooling
neutron star (Vasisht et al. 1997).  Less than 24\% of the ASCA flux is
pulsed.

\subsubsection{1E 161348--5055 (RCW 103)}

Einstein HRI observations first revealed 1E 161348--5055 as an
unresolved X--ray source within RCW 103 (Tuohy \& Garmire 1980).  With
the object lying very close to the centre of the SNR in a minimum of the
diffuse shell emission, an association between the two seemed likely,
but searches for optical or radio counterparts have not been successful.
Kaspi et al. (1996) provide 400 MHz and 1.5 GHz pulsed flux density
limits of 1.5 mJy and 0.1 mJy respectively.  Recently, the point source
has been observed with ASCA and separated spectrally from the bright but
softer SNR shell emission (Gotthelf et al. 1997).  The point source flux
is probably constant on long time scales, within the errors presented by
matching an ill-defined spectrum across several instruments.  The age of
the remnant is estimated to be 1,000--3,000 years, and the supernova
responsible for RCW 103 might have been a guest star reported in
134 B.C. (Wang et al. 1986).

\subsubsection{RXJ2020.2+4026 (G078.2+2.1)}

The \gam--ray source 2EG J2020+4026 (2CG078) has been linked with the
G078.2+2.1 SNR by several authors, most recently Sturner \& Dermer
(1995) and Esposito et al. (1996).  During a detailed study of the
\gam--ray source, Brazier et al. (1996) noted the existence of a single
unresolved X--ray source, RXJ2020.2+4026, at the centre of the remnant.
The X--ray flux is steady and no likely optical or radio counterparts
were found in subsequent searches, leaving the possibility that the
X--ray and \gam--ray fluxes were from a pulsar in the G078.2+2.1 SNR.
Like RXJ0007.0+7302, the proposed pulsar is Vela-like, with a small
X--ray flux relative to the \gam--ray flux.

\subsection{Isolated candidates}

\subsubsection{Geminga}

The discovery of 4.2 Hz pulsations in the enigmatic Geminga,
first in X--rays (Halpern \& Holt 1992) and then in \gam--rays (Bertsch et
al. 1992), confirmed this as a pulsar, albeit one with a radio
luminosity very much lower than any known radio pulsar.  Recent reports
that Geminga has finally been seen as a radio pulsar at very low
frequencies (Kuzmin \& Losovsky 1997, Malofeev \& Malov 1997) do not affect
the limits on its luminosity at higher frequencies, which we use here
for comparison with the population of radio pulsars.

The lack of absorption of Geminga's X--ray flux and its high \gam--ray
flux mean that Geminga must be nearby, within the approximate range $
100 < d < 400\,$pc (Bertsch et al. 1992, Halpern \& Ruderman 1993).  A
recent, marginal detection of parallax in Hubble Space Telescope
observations gives a distance of 120--220 pc (Caraveo et al. 1996), in
good agreement with the earlier estimates.  The best parallax value of
160 pc will be adopted in this paper  

Knowledge of Geminga's spin parameters and distance enables a direct
comparison with radio pulsars.  In particular, the spin-down power and
age can be derived from the rotation period $P$ and its first derivative
$\dot P$ in exactly the same way as for other pulsars.

\subsubsection{MS 0317--6647}

Among the list of candidate neutron stars, MS 0317--6647 is the only one
to show signs of long-term variability.  It is an unusual source, with
no optical counterpart and a hard, featureless X--ray spectrum that is not well
described by simple models (Petre et al. 1994).  While it may well be a
luminous X--ray binary in the spiral galaxy NGC 1313, Stocke et al.
(1995) discuss the possibility that it is an old neutron star in our own
galaxy.  The variability argues against a rotation-powered or cooling
neutron star, leaving the option that this is an object accreting from
the interstellar medium (ISM).  No specific radio flux limits are
available, and this object is not included further in this paper.

\subsubsection{RXJ1856.5--3754}

Walter et al. (1996) announced the discovery that this X--ray source was
an old neutron star at a distance of about 100 pc.  The source, first
detected in the Einstein slew survey, is very bright, yet no 
counterpart was found, and the source flux did not vary.  The
proposed distance to the object was determined from the X--ray spectrum
of the object: since the source is in the direction of a molecular cloud
at $\sim 120$ pc, the low $N_H$ measured from the X--ray spectrum implied
that the source lay in front of the cloud.  Walter et al. concluded that
the object was a nearby, old neutron star, perhaps powering its X--ray
emission through accretion from the interstellar medium.

Reassessments of RXJ1856.5--3754 have been published by Neuh\"auser et
al. (1997) and Campana et al. (1997).  Both groups find that their more
conservative error circles for the X--ray position still exclude all
bright optical sources, and they agree with the previous conclusion that
the object is probably a neutron star.  However, it has not been
possible to distinguish between a ``middle-aged'' pulsar like Geminga
and an old neutron star accreting from the ISM. Using the revised source
position given by Neuh\"auser et al., Walter \& Matthews (1997) have now
identified a faint optical counterpart at magnitude 25.6, consistent
with a neutron star.  The distance to the object may be larger than
claimed by Walter et al. (1996).  The X--ray spectrum is well modelled
by black body radiation from the surface of a neutron star at 100--170
pc. However, the X--ray spectra of radio pulsars do not give reliable
distances under the same assumptions.  Observations of the parallax
and/or proper motion of RXJ1856.5--3754 will help to resolve its nature
and distance.  For the rest of this paper we will assume that it is at a
distance of 120 pc.

\begintable*{2}
\tabletext{
{\bf Table 2.} Offsets of pulsar candidates from SNR centres, expressed
as fractions $\beta$ of the SNR radius.  The implied velocities, using
the distance and age estimates in Table 1, are also given.  Supernova sizes
and centres are from Green's SNR catalogue (1996) and from Petre et al. 
(1996).  The size of the 3C58 remnant is for the plerion alone.}
\halign {$\hbox{#}\hfil$ & $\hfil\hbox{#}\quad$ & $\hfil#$\quad\quad\quad & $\hfil#$\quad\quad&$\hfil#$\quad\quad&$\hfil#$\cr
\noalign{\hrule\vskip 0.03in}\cr
\noalign{\hrule\vskip 0.08in}
Compact source & SNR & \hbox{SNR radius}\hskip -20pt & \hbox{Offset}\hfil  & \beta\hfil &  v \hfil     \cr
{}	       & {}  & \hbox{(arcmin)}\hfil\hskip -20pt   & \hbox{(arcmin)}          &       & (\hbox{km}\,\hbox{s}^{-1}) \cr
RXJ0007.0+7302  &\hbox{CTA 1}       &45   &15\quad     &0.33   &730\quad    \cr
RXJ0201.8+6435	&\hbox{3C58}        &3    &?\quad      &?      &{?}\quad\cr
1E 0820--4247	&\hbox{Pup A}       &25   &6\quad      &0.25   &930\quad  \cr
1E 1207.4--5209	&\hbox{G296.5+10.0} &33   &3\quad      &0.10   &200\quad  \cr
1E 161348--5055	&\hbox{RCW103}      &10   &0.5\quad    &0.05   &220\quad  \cr
RXJ2020.2+4026	&\hbox{G078.2+2.1}    &60   &1.5\quad    &0.03   &60\quad   \cr
\noalign{\vskip 0.03in\hrule\vskip 0.08in}
}
\endtable

\vskip 15pt

From the ten objects listed in Table 1 we conservatively accept the
eight with quantitative radio limits as good neutron star candidates,
six in SNRs and two isolated. The strong case for physical association
between the first group of X--ray sources and the supernova remnants is
seen more clearly if we consider the distances from the point sources to
the SNR centres and the transverse velocities that these distances
imply.  With the exclusion of 3C58, which has no known SNR shell, these are listed
in Table 2.  The velocities are entirely consistent with the median of
460 km$\,$s$^{-1}$ given by this method for radio pulsars in SNRs (Frail
et al. 1994) and the median of 300 km$\,$s$^{-1}$ seen in the general
population of radio pulsars (Lorimer et al. 1997a).  Also listed in
Table 2 is the angular distance $\beta$ between the pulsar and the SNR
centre, expressed as a fraction of the SNR radius.  The values of
$\beta$ here are {\it smaller} than in most suggested associations
between radio pulsars and supernova remnants (Kaspi 1996 \& references
therein).  It is improbable that so many unusual X--ray objects would be
found near the centres of young SNRs unless there is an association. In
addition, the plerions in CTA 1 and 3C58 point very strongly to active
sources of relativistic particles in these SNRs.  The lack of a plerion
in the other SNRs, however, does not imply the reverse (Bhattacharya
1990).

For the six candidates in supernova remnants, there is therefore strong
evidence that they are neutron stars.  Of the two isolated candidates,
Geminga is confirmed as a pulsar and RX J1856.5--3754 has properties
entirely consistent with the proposition that it too is a neutron star.
For the remainder of this paper, it will be assumed that all eight
candidates are neutron stars.

\beginfigure*{1} \epsfxsize=14.0cm\epsfbox{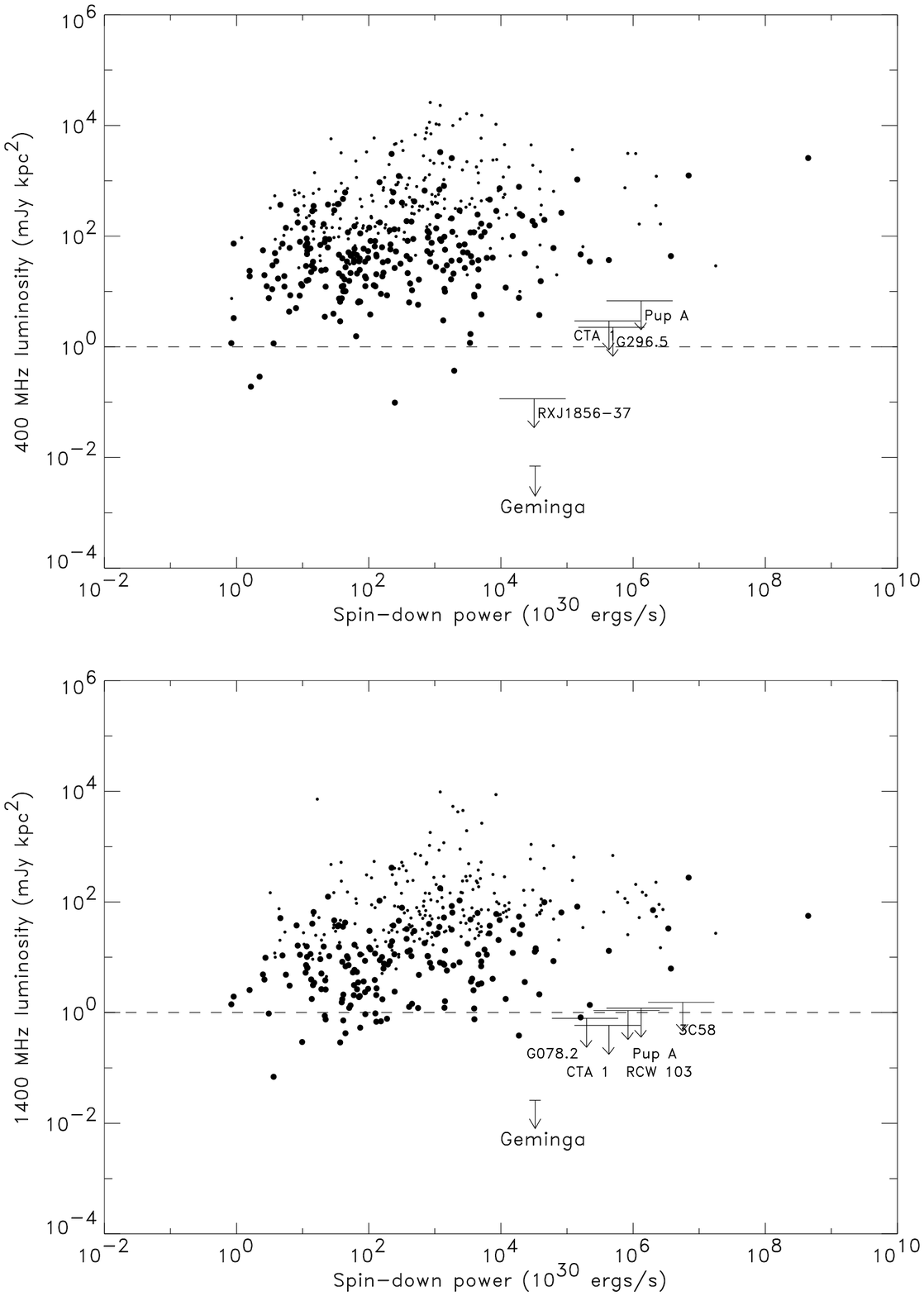}

\caption{{\bf Figure 1.} The luminosities of radio pulsars at 
400 MHz (top) and 1.4 GHz (bottom) against spin-down power.  Larger
symbols indicate pulsars at distances of less than 3.5 kpc.
Restrictive upper limits are included for Geminga and the X--ray-selected 
candidate neutron stars selected from Table 1.  Dashed horizontal
lines illustrate a luminosity of $1\,\rlum$.}
\endfigure

\beginfigure*{2} \epsfxsize=14.0cm\epsfbox{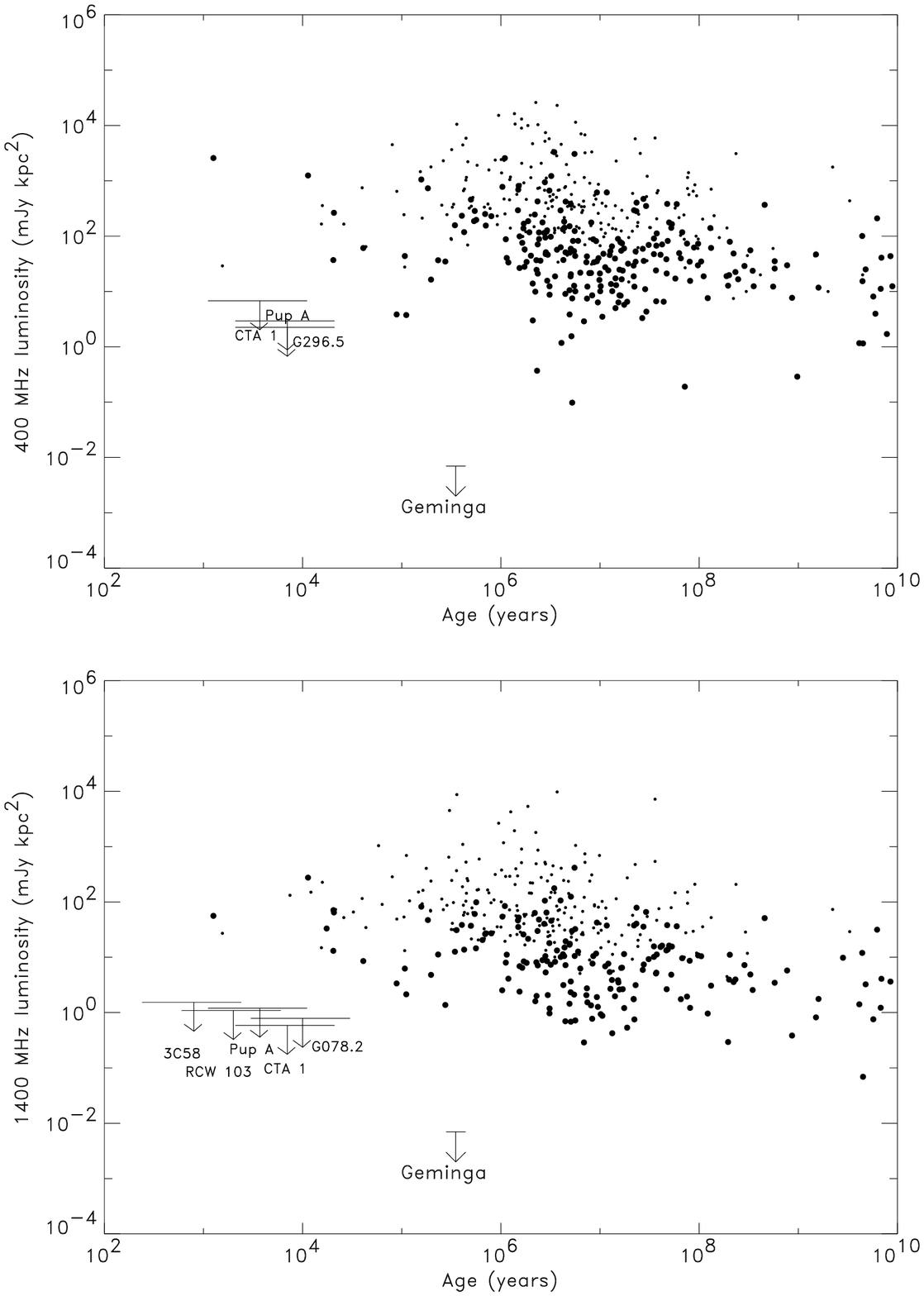}

\caption{{\bf Figure 2.} The luminosities of radio pulsars and candidate
neutron stars at 400 MHz (top) and 1.4 GHz (bottom) against pulsar or
SNR age, where pulsar ages are calculated from $\tau = p/2\dot p$ and
SNR ages are listed in Table 1.  Larger symbols show the positions of
pulsars with distances of less than 3.5 kpc.}  
\endfigure

\section{Comparing radio pulsars with X--ray-selected neutron star candidates}

None of the eight accepted objects has been detected as a radio
source.  Could any of them be a normal radio pulsar?  In Fig.~1 the
radio luminosities, assuming 1 steradian beaming, of radio pulsars
listed in the Princeton pulsar catalogue (Taylor et al. 1993) are
plotted against spin-down power.  Larger symbols indicate pulsars at
distances of less than 3.5 kpc, directly comparable with the X--ray
pulsar candidates.  The axes demonstrate the rise towards a higher
median luminosity, $\sim 300\,\rlum$ at 400 MHz, $\sim 100\,\rlum$ at
1400 MHz, for energetic pulsars (e.g. Lyne et al. 1985, Taylor \&
Stinebring 1986, Tauris \& Manchester 1997) and the sensitivity limits
of radio searches.  Using Table 1, the upper limits for the seven pulsar
candidates and Geminga have also been added.  Each of the candidate
pulsars is shown with an order of magnitude uncertainty in the spin-down
power: this is intended to be illustrative and does not indicate a
formal uncertainty.  Note that for the candidates, {\it both} axes scale
with the square of distance.

An alternative way to view these data is in terms of pulsar/candidate
age, as shown in Figure 2.  In this figure, the ages for the candidate
neutron stars are taken to be those of the host SNR, again giving a
dependence on distance.  Increasing the distances in both figures, for
example, would tend to make the radio limits weaker and the candidate
pulsars older but also more powerful.  Even taking reasonable degrees of
uncertainty into account, very large increases in distance would be
required to bring the radio luminosities of the candidate pulsars into
the realm of known radio pulsars.  There are four possible explanations
for their non-detection as radio sources:

\beginlist 
\item{1.} They have intrinsically small radio luminosities or very 
steep radio spectra

\item{2.} The radio beams are directed away from Earth

\item{3.  They have unusual spin parameters untested by current pulsar
searches.}

\item{4.  They are not pulsars.}  
\endlist 

Although we have accepted all eight candidates as
neutron stars, it will of course remain possible that one or more of
the candidates is not a rotation-powered pulsar until each of them has
been identified conclusively.  Weak accretion has been suggested as a
possible power source in RXJ1856.5--3754 and 1E~1207.4--5209.  The
accretion models fall into two types: old neutron stars accreting from
the interstellar medium and `anomalous X--ray pulsars' (van Paradijs et
al. 1995) with periods of a few seconds and large X--ray luminosities
relative to the power available from their spin down.  The latter
objects are thought to be isolated neutron stars powering their
X--radiation by accretion and to result from a high mass X--ray binary
(HMXB) which underwent a common envelope phase (van Paradijs et
al. 1995, Ghosh et al. 1997).  The distances and ages of the four
examples known give a total of only 20--50 such pulsars in the Galaxy, a
very low number appropriate for their exotic history.  Therefore,
although these objects are difficult to distinguish from other pulsars
without the detection of pulsations and are radio-quiet, it is unlikely
that they explain all the candidate neutron stars in Table 1.  Old neutron
stars producing detectable X--ray fluxes by accretion from the
interstellar medium must also be rare because deep searches have failed
to find any (Manning et al. 1996, Danner 1997a, 1997b).  This mechanism
is a possible explanation for only one of the candidate neutron stars,
RXJ1856.5--3754.

Empirically, the underlying luminosity function of nearby radio pulsars
flattens below 20 mJy kpc$^2$ (Lyne et al. 1997), giving no evidence for
substantial numbers of low luminosity pulsars.  In addition, proposing
that the candidate pulsars have low radio luminosities or steep spectra
leads to a conflict with such population constraints, unless
there is some mechanism that can suppress the radio emission from young
pulsars but permits high energy radiation and the formation of plerions.
This is a particular problem for the candidates in young SNRs, since we
would expect large numbers of similar objects at greater ages.

It has previously been proposed that pulsars with extreme parameters,
such as unusually long periods or high magnetic fields, might have radio
luminosities low enough to escape detection.  For example, slow ($P >
0.3\,$s) pulsars with low magnetic fields might be `injected' into the
population, forming a sub-population of low luminosity pulsars (Narayan
1987).  However, there is no compelling observational evidence for
injected pulsars.  A number of authors have now shown that the results
of radio pulsar surveys and the small number of pulsar/SNR associations
can be explained at least as well by a simple, single population 
as by multi-population models (Bhattacharya et al. 1992, Frail \& Moffett
1993, Lorimer et al. 1993, Gaensler \& Johnston 1995, Lorimer et
al. 1997b).

\vskip 10pt

The simplest explanation for the candidate radio-quiet pulsars is that
they are radio pulsars whose beams do not sweep past the Earth.  This
requires no new population of neutron stars and can accommodate all of
the candidates and Geminga.  However, it has a number of implications
for pulsar beaming and pulsar statistics, discussed below.

\subsection{Beaming}

Pulsars are complex X--ray sources.  The very young, Crab-like radio
pulsars are visible in soft X--rays as non-thermal, highly pulsed,
compact sources surrounded by a bright nebula of synchrotron X--rays.
Older pulsars, up to a few $\times 10^5$ years old, are
generally much weaker point sources and usually lack a synchrotron
nebula.  In these pulsars, the non-thermal pulses become dominant only
above a few keV, while the X--rays in the ROSAT (0.1 -- 2.4 keV) band are
thermal radiation from the $10^6\,$K neutron star surface, 
representing the cooling of the hot stellar interior and bombarding
of the polar caps by fast particles from the magnetosphere.
The polar caps of the neutron star are hotter than the rest of the
surface and are detected as pulses, but because the radiation is bent
gravitationally, the pulses are spread out to give a low degree of
modulation even when radiation is entirely from the polar caps (Zavlin
et al. 1995, Page 1995, Yancopoulos et al. 1994).  Typically less than
20\% of the keV flux in detected pulsars forms the characteristically
broad, smooth pulse, in agreement with the predictions.  We will
therefore assume that this thermal radiation is spread by gravitational
bending to be visible from all directions.  This means that our ability
to detect a neutron star from its soft X--ray emission is unaffected by
beaming.

Thermal emission appears to be responsible for the X--rays from the
candidate neutron stars, given their low luminosities, soft spectra and
the lack of pulsations below 2 keV.  Where the spectra have been
modelled, they have been described as blackbody, although not all of
them are well constrained.

The beaming fraction of radio pulsars is not well known.  It is
generally accepted that the average beaming fraction for the whole
population is around 20\% (e.g. Lyne \& Manchester 1988).  Tauris \&
Manchester (1997) find that it is only 10\% but anti-correlates with
age, giving a much higher beaming fraction for young pulsars. Some
authors (e.g. Narayan 1987) find a beaming fraction approaching 100\%
for young pulsars, boosting their case for injection of long-period
pulsars into the population.  Frail \& Moffett (1993) derived a value of
$61 \pm 13\%$ from a deep search for (young) radio pulsars in plerions.

We can provide an independent measure of the beaming fraction by
considering the pulsars which have been detected in X-rays but are
radio-quiet.  Taking just the pulsar candidates with quantitative radio
flux limits, the six in SNRs have ages less than 20,000 yr and distances
smaller than 3.5 kpc.  Six known radio pulsars also satisfy these criteria:
PSRs B0531+21, B0833--45, B1706--44 and B1046--58, all of which have
been detected at high energies, and PSRs B1737--30 and B1853+01, which
have not.  These latter two pulsars are relatively low down on the $\dot
E/d^2$ ranking, probably explaining why they have not been detected.
The number of radio-quiet pulsars relative to the total yields a radio
beaming fraction of $\sim 50\%$ for young pulsars, assuming that the
sample is largely complete.  We argue below that this is the case.  The
radio-quiet pulsars are clearly inconsistent with beaming fractions close to
100\%.

\subsubsection{Geminga-like \gam--ray pulsars}

The presence of pulsed \gam--rays from a radio-quiet pulsar allows us to
constrain the emission geometry of the radio and high energy beams.  At
present \gam--ray pulses have been identified only in Geminga, although
the candidate pulsars RXJ0007.0+7302 and RXJ2020.2+4026 are also
coincident with \gam--ray sources.  The beaming explanation for
radio-quiet neutron stars therefore demands a model in which it is
possible to see the hard X--/\gam--ray pulses without intersecting the
radio beam.

Current data on the high energy emission from pulsars is limited by the
sensitivity of available instruments.  In \gam--rays, just six radio pulsars
have been detected as pulsed sources (Thompson et al. 1994, Ramanamurthy
et al. 1995, Carrami\~nana et al. 1995).  The range of pulse shapes,
from a single broad hump to two widely separated, sharp peaks connected
by a saddle, can be explained in terms of different lines of sight
across a single, edge-brightened beam (Daugherty \& Harding 1996, Romani
\& Yadigaroglu 1995).  The wide pulses imply a broad beam unless there
are only small offsets between the observer, the beam axis and the
neutron star spin axis.  Wider beams are generally preferable because
they do not require such a specific geometry and can explain why all of
the radio pulsars with highest $\dot E/d^2$ have been detected.

In order to explain radio-quiet pulsars by beaming, the high energy and
radio beams cannot be generated in the same location in the neutron star
magnetosphere.  One possibility (e.g. Romani \& Yadigaroglu 1995) is
that the high energy beams are produced near the light cylinder
(co-rotation radius) while the radio is beamed along the magnetic axis.
Efficient production of high energy photons in this model depends on a
high inclination of the magnetic axis (Romani 1996) and a pulsar with
low magnetic inclination or viewed from near the spin axis would not be
visible at high energies.  Approximately two-thirds of the pulsars
visible as EGRET \gam--ray sources would be radio-quiet in this model
(Yadigaroglu \& Romani 1995), although a larger radio beaming fraction
and allowance for deep, targeted radio pulsar searches will decrease
this figure.  Several groups are working on unidentified \gam--ray
sources to look for radio-quiet pulsars.

Uncertainty over the inclinations measured from radio polarisation (Lyne
\& Manchester 1988, Rankin 1990, Manchester 1996) fuels an on-going
debate between the above model and an alternative set of models with
exactly the opposite geometric requirements (Daugherty \& Harding 1996,
Sturner \& Dermer 1995).  These `polar cap' models have been successful
in explaining the spectra and \gam--ray luminosities of radio pulsars,
but they require that the pulsar magnetic and spin axes are
approximately aligned in order for the emission from a single magnetic
pole to produce the broad observed pulses. The most recent simulations
(Daugherty \& Harding 1996) have more relaxed geometries than earlier
models.  If polar cap models are to explain radio-quiet pulsars by
beaming, then the radio beams come either from a separate region of the
magnetosphere or are internal to the hollow \gam--ray cone.  The radio
pulses of \gam--ray pulsars are (so far) always {\it outside} the
\gam--ray pulse, inconsistent with an internal radio beam (Daugherty \&
Harding 1996).  In both outer gap and polar cap models, therefore, the
radio and high energy beams of radio-quiet pulsars must be generated in
different parts of the magnetosphere.

\subsection{The pulsar birth rate}

Pulsar birth rates are usually derived from observations and models of
the radio population alone.  In this section we use the X-ray observations of
young neutron stars, usually neglected in such calculations, to
constrain the local pulsar birth rate {\it independent} of radio beaming
and luminosity laws.

As we listed in Section 3.1, 10 neutron stars with ages less than 20,000
years and distances below 3.5 kpc have been detected in X--rays.  This
implies a birth rate of 13 Myr$^{-1}\,$kpc$^{-2}$, if pulsars are born
close to the Galactic disk and the X-rays are not beamed.  In order to
extrapolate this to the whole Galaxy, we assume that the radial
distribution of pulsars is a Gaussian with a radial scale length of 5
kpc (e.g. Lorimer et al. 1993), which gives a Galactic neutron star birth
rate of 1 every 110 years.  Adding two radio pulsars that have not yet
been detected in X-rays raises this figure to 1 every 91 years.  Further
allowance for incompleteness in the X-ray detections can only increase
the birth rate further.

The frequency of supernova explosions and the birthrates of SNRs and
pulsars have been the subject of discussion since the late 1960's.
Recent calculations give a total (Type I plus Type II) Galactic
supernova rate of 1 every 40 years (Tammann et al. 1994), similar to the
rate of one Type II every 50--170 years derived from extragalactic supernova
searches (Cappellaro et al. 1997).  A recent estimate for the birth rate
of radio pulsars is 1 every 60 -- 330 years (Lyne et al. 1997).
Gaensler \& Johnston (1995) found that a birth rate of one every 85
years gave an excellent match between observed and modelled SNR/pulsar
associations.

The birth rate cannot be a factor of 2 higher than we have derived from
the X-ray neutron star population, or it will be in conflict with the
(independently derived) supernova rate.  It is also unlikely to be much lower
than our estimate because our sample is not complete.  The birth rates
for neutron stars, Type II supernovae and radio pulsars are therefore
similar.  We conclude that (a) probably all young neutron stars are
radio pulsars, (b) more young pulsars are visible as X-ray sources than
as radio pulsars, and (c) most of the young, nearby pulsars have already
been discovered.

Lastly, our result supports the conclusion of Gaensler \& Johnston (1995)
that the small number of known pulsar/SNR associations is a consequence
of pulsar beaming and luminosity, not a dearth of radio pulsars.

\section{Conclusions}

This paper has clarified the evidence for radio-quiet pulsars and the
implications of such objects.  We have listed six clear, unresolved X--ray
sources in supernova remnants, with quantitative radio flux limits and
high X--ray to optical flux ratios that rule out nearly all types of
X--ray source. Most of them are the only X--ray point source within their
SNR and they are all close to their SNR centres, making a strong case that
they are stellar remnants associated with the SNRs.  Their transverse
velocities are consistent with the velocities observed in radio pulsars
(Frail et al. 1994, Lorimer et al. 1997a).  We find that it is simplest
to explain all of these objects, plus two further objects without SNRs, as
neutron stars.

The candidate neutron stars have lower radio fluxes than would be
expected from known radio pulsars of equivalent age or spin-down power.
Reasons for this might include extreme spin parameters (e.g because of
large magnetic fields) or truly low radio luminosities.  However, these
are not necessary to explain the sources or justified by other empirical
evidence.  The low radio luminosities are most simply accommodated in a
geometric explanation, in which the radio emission is not favourably
beamed whereas the soft X--rays are dominated by thermal emission from
the neutron star surface and are visible from all directions.  The
relative numbers of radio pulsars and X--ray pulsar candidates in SNRs
gives a crude estimate of $\sim 50\%$ for the radio beaming fraction.

Above $\sim 2$ keV, non-thermal emission from the magnetosphere becomes
dominant in known pulsars.  The presence of high energy radiation
without radio pulses implies different emission sites for the two ends
of the spectrum.  The candidate pulsars in CTA 1 and G078.2+2.1 coincide
with \gam--ray sources and searches for pulsations in the high energy
fluxes should be pursued.  1E~1207.4--5209, the candidate neutron star in
G296.5+10.0, has not shown any evidence for magnetospheric emission.
This could mean that the object is a cooling neutron star with only weak
magnetospheric activity (Vasisht et al. 1997), but it could also be a
pulsar in which both the radio and high energy beams are directed away
from the Earth.  The object's rotation frequency may still be discovered
from low-level modulations of the thermal soft X--rays.

We have also used the assumption of quasi-isotropic X--ray emission to
estimate the neutron star birth rate, which we find to be at least 13
Myr$^{-1}\,$kpc$^{-2}$ in the neighbourhood of the Sun.  The total
galactic birth rate is therefore at least 1 neutron star every $\sim 90$
years, close to the derived rate of Type II supernovae (Cappellaro et
al. 1997).  We conclude that neutron stars are a frequent outcome of
supernovae, that probably all neutron stars are born as radio pulsar and
that most young, nearby pulsars have already been discovered.  This is
further support for our result that radio-quiet pulsars are best
explained as unfavourably beamed radio pulsars.

\section{Acknowledgments} 

We thank Bryan Gaensler for his constructive reading of this paper.
KTSB thanks the STARLINK project for provision of computing facilities
and PPARC for financial support.

\bye